\documentclass[prb,nofootinbib,twocolumn,superscriptaddress]{revtex4} 

\usepackage{graphicx}
\usepackage{dcolumn}
\usepackage{bm}
\usepackage{threeparttable}
\usepackage{times}
\usepackage{mathptmx}
\usepackage{lscape}
\usepackage{natbib}
\usepackage{amsmath}
\usepackage{amssymb}
\usepackage{braket}
\usepackage{comment}
\usepackage{color}

\def\degree{\kern-.2em\r{}\kern-.3em}

\begin{document}


\title{  Configurational Geometry Bridges Equilibrium Structure Information \\from a Single to Multiple Compositions   }

\author{Shouno Ohta}
\affiliation{
Department of Materials Science and Engineering,  Kyoto University, Sakyo, Kyoto 606-8501, Japan\\
}%

\author{Ryogo Miyake}
\affiliation{
Department of Materials Science and Engineering,  Kyoto University, Sakyo, Kyoto 606-8501, Japan\\
}%

\author{Koretaka Yuge}
\affiliation{
Department of Materials Science and Engineering,  Kyoto University, Sakyo, Kyoto 606-8501, Japan\\
}%

\begin{abstract}
{ For classical discrete systems under constant composition, a set of microscopic state dominantly contributing to thermodynamically equilibrium structure should depend on temperature and energy through Boltzmann factor, $\exp\left( -\beta E \right)$. Despite this fact, our recent study find a set of special microscopic state that can characterize equilibrium properties, where these structures can be know a priori without any thermodynamic information. Here, for binary system, we extend the theoretical approach to develop a new formulation, where the special microscopic states at a given, \textit{single} composition can characterize equilibrium structure over \textit{whole} composition. We demonstrate the validity of the proposed formulation by comparing with results by conventional thermodynamic simulaton. The results strongly indicate that most information about composition- and temperature-dependence of thermodynamically equilibrium structure for disordered state on multiple compositions can be concentrated to a set of special microscopic state on any single composition.
  }
\end{abstract}


\maketitle

\section{Introduction}
For classical discrete systems typically refered to substitutional crystalline solids under \textit{constant} composition, $r$-th component of structure under coordination of $\left\{ Q_{1},\cdots,Q_{f} \right\}$ for thermodynamically equilibrium state can be given by the following canonical average:
\begin{eqnarray}
\label{eq:can}
\Braket{Q_{r}}_{Z} = Z^{-1} \sum_{d} Q_{r}^{\left( d \right)} \exp\left( -\beta E^{\left( d \right)} \right),
\end{eqnarray}
where $Z$ denotes partition function, $\beta$ inverse temperature, and summation is taken over all possible microscopic states on configuration space. 
From Eq.~\eqref{eq:can}, we can clearly see that a set of microscopic state dominantly contributing to the l.h.s. of $\Braket{Q_{r}}_{Z}$ should depend on 
temperature as wll as on energy (i.e., many-body interaction). Since a number of possible microscopic states astronomically increase with increase of system size, 
a variety of theoretical approaches has been amply developed to efficiently sample important states including Metropolis algorism, entropic sampling and Wang-Landau method.\cite{mc1,mc2,mc3,wl} 
Despite these facts, we recently derive that Eq.~\eqref{eq:can} can be significantly simplified to the sum of a set of special microscopic states:\cite{em1,em2,em0}
\begin{eqnarray}
\label{eq:emrs}
\Braket{Q_{r}}_{Z} \simeq \Braket{Q_{r}} -  C_{r}\beta E_{r} + \frac{\beta^{2}}{2}\sum_{i=1}^{g} \pm E_{r_{i}}^{2},
\end{eqnarray}
where $\Braket{\quad}$ denotes taking arithmetic average for configurational density of states (CDOS) \textit{before} applying many-body interaction to the system, $C_{r}$ represents constant depending only CDOS geometry, and $E_{r}$ and $E_{r_{i}}$s are energy for the special microscopic states, whose structures can be known \textit{a priori} without any thermodynamic information, depending only configurational geometry (i.e., information about the CDOS). We call these special microscopic states as PS (projection state) and PS2s, respectively.
Although Eq.~\eqref{eq:emrs} can hold for systems on any lattice and on number of components, its application is restricted to a given single composition, i.e., information about 
energy for PS and PS2 cannot be used for predicting $\Braket{Q_{r}}_{Z}$ for other compositions. 
We here extend our theoretical approach, where information about PS energies on a \textit{single} composition can predict canonical average of structure for any given compositions. The details and derived modified formulations are shown below.

\section{Derivation and Discussions}
Before derivating modified formulation of Eq.~\eqref{eq:emrs} for whole composition, we first briefly see the characteristic structure for PS and PS2. 
For simplicity, hereinafter we describe structure measured from 1-order moments, $\Braket{Q_{r}}$s.
In the present study, as we see later, we confine ourselves to systems under pair correlations: Then, $m$-th coordination corresponds to $m$-th nearest-neighbor (m-NN) pair correlation. Intuitively, structure of PS along $r$-th coordination is constructed from information about multivariate 2-order moments $\mu_{2}$ on a given composition. Its $j$-th component is given by
\begin{eqnarray}
\label{eq:2mom}
Q_{j}^{\left( E_{r} \right)} \simeq \sqrt{\frac{2}{\pi }}\Braket{Q_{r}}_{2}^{-1} \mu_{2}^{\left( r,j \right)}, 
\end{eqnarray}
where $\Braket{\quad}_{2}$ denotes taking standard deviation for CDOS, and  $\mu_{2}^{\left( r,j \right)}$ represents 2-order moment for CDOS projected onto $\left( Q_{r}, Q_{j} \right)$ space. In this case, energy for PS can be expressed as
\begin{eqnarray}
E_{r} = \sum_{j=1}^{f} \Braket{E|Q_{j}} Q_{j}^{\left( E_{r} \right)},
\end{eqnarray}
where $\Braket{a|b}$ denotes inner product, i.e., trace over configuration space. 
Meanwhile, $j$-th component of $i$-th structure of PS2 along $r$-th coordination is given by
\begin{eqnarray}
\label{eq:3mom}
Q_{j}^{\left( E_{r_{i}} \right)}  = \lambda_{i}^{\frac{1}{2}} U_{ij},
\end{eqnarray}
where $\lambda_{i}$ and $U_{ij}$ are $i$-th singular value and $\left( i,j \right)$ component of left singular vector for real symmetric $f\times f$ 3-order moment matrix $\mathbf{A}$ defined as 
\begin{eqnarray}
\label{eq:a}
A_{pq} = \Braket{Q_{r}Q_{p}Q_{q}} =\mu_{3}^{\left( r,p,q \right)}.
\end{eqnarray}
The important points in Eqs.~\eqref{eq:2mom} and~\eqref{eq:3mom} are that individual moments should be taken for CDOS under a given single composition, which cannot be straightforwardly applied for predicting canonical average for other compositions. 

In order to extend such applicable limitation, we start from exact formulation for multivariate lower order moments for binary system under constant composition derived by our recent study. From the study, 2-order moment between $r$- and $j$-th pair coordination is given by
\begin{widetext}
\begin{eqnarray}
\mu_{2}^{\left( r,j \right)} = \frac{16 z \left( z N^{2} - N + 1 \right)}{ \left( N-1 \right)^{2} \left( N-2 \right)  \left( N-3 \right) D_{j} } \left\{ -2D_{j} + \left( N-1 \right)\left[ r=j \right] \right\},
\end{eqnarray}
\end{widetext}
and 3-order moment for $r$-, $j$- and $k$-th pair coordination is given by
\begin{widetext}
\begin{eqnarray}
\mu_{3}^{\left( r,j,k \right)} = \frac{ 64 z \left( z N^{2} - N + 1 \right)  }{ N \left( N-1 \right)^{3} \left( N-2 \right) \left( N-3 \right) \left( N-4 \right) \left( N-5 \right) D_{r} D_{j}  D_{k}} \left\{   \left( N-1 \right)^{2} \left( zN^{2} - 2N + 4 \right) c_{rjk}  - 8N \left( zN^{2} + \left( 5z - 3 \right) N + 3  \right) D_{r} D_{j}  D_{k} \right. \nonumber  \\ 
\left.  + \left( N-1 \right)^{2} \left( \left( 1-4z \right)N^{2} -N  + 4  \right) D_{r} \left[ r=j=k \right]  + 2\left( N-1 \right)\left( \left( -1+6z \right)N^{2} - 3N+4 \right)\left( D_{k}D_{r}\left[ r=j \right] + D_{r}D_{j}\left[ j=k \right] + D_{j}D_{k}\left[ k=r \right] \right)  \right\}. \nonumber \\
\quad
\end{eqnarray}
\end{widetext}
Here, $N$ denotes number of lattice points in the system, 
\begin{eqnarray}
z=x\left( 1-x \right)
\end{eqnarray} 
in A$_{x}$B$_{\left( 1-x \right)}$ binary composition, $D_{r}$ represents number of $r$-NN pair per site, $c_{rjk}$ number of possible closed path consisting of $r$, $j$ and $k$-NN pair (e.g., $c_{111} = 6T_{111}$, $c_{112}= 2T_{112}$ and $c_{123}=T_{123}$ where $T_{ijk}$ denotes number of triples per site), and $\left[ \quad \right]$ corresponds to Iverson bracket defined as
\begin{eqnarray}
\left[ P \right] = \left\{ \begin{array}{ll}
1 & \left( P \;\mathrm{is \;true} \right) \\
0 & \left( otherwise \right).
\end{array} \right.
\end{eqnarray}
Then, our strategy is to see the moments at thermodynamic limit of $N\to\infty$:
\begin{eqnarray}
\label{eq:moms}
\lim_{N\to\infty} \mu_{2}^{\left( r,j \right)} &=& \frac{16z^{2}\delta_{rj}}{N D_{j} } \nonumber \\
\lim_{N\to\infty} \mu_{3}^{\left( r,j,k \right)} &=& \frac{64z^{2}}{N^{2}D_{r} D_{j} D_{k} } \left\{ \left( 1-4z \right) D_{r} \left[ r=j=k \right] + zc_{rjk}  \right\}. \nonumber \\
\end{eqnarray}
From above discussions and equations, we can rewrite canonical average of Eq.~\eqref{eq:emrs} as 
\begin{widetext}
\begin{eqnarray}
\label{eq:mod}
&&\Braket{Q_{r}}_{Z} \simeq \Braket{Q_{r}} - \beta  \frac{16z^{2}}{N D_{r} } \Braket{E|Q_{r}} + \frac{\beta^{2}}{2} \left\{ 64z^{2}\left( 1-4z \right) \frac{1}{\left( ND_{r} \right)^{2}   } \Braket{E|Q_{r}}^{2}  + \sum_{j=1}^{f}\sum_{k=1}^{f} 64z^{3} \frac{c_{rjk}}{D_{r}D_{j}D_{k}N^{2} }\Braket{E|Q_{j}}\Braket{E|Q_{k}}\right\} \nonumber \\
&&= \Braket{Q_{r}} - 16z^{2}\beta \frac{\Braket{E|Q_{r}}}{ND_{r} } + 32z^{2}\left( 1-4z \right)\beta^{2}   \left( \frac{\Braket{E|Q_{r}}}{ND_{r} } \right)^{2} + 32z^{3}\beta^{2} \sum_{i=1}^{h} \omega_{i} \left\{  \sum_{m=1}^{f} \Braket{E|Q_{m}}\left( \gamma_{i}^{\frac{1}{2}} V_{im} \right)  \right\}^{2},
\end{eqnarray}
\end{widetext}
where $\gamma_{i}$ denotes $i$-th singular value, $\omega_{i}$ takes +1 (-1) when corresponding eigenvalue takes positive (negative) sign, and $V_{im}$ denotes $\left( i,m \right)$ component of l.h.s singular matrix for real symmeric matrix $\mathbf{B}$ with $\mathrm{rank} \:\mathbf{B} = h$:
\begin{eqnarray}
B_{pq} = \frac{c_{rpq}}{D_{r}D_{p}D_{q} N^{2} }. 
\end{eqnarray}
From Eq.~\eqref{eq:mod}, if the $0$-th microscopic state has structure of
\begin{eqnarray}
Q_{k=r}^{\left( 0 \right)} &=& \frac{1}{ND_{r}} \nonumber \\
\quad Q_{k\neq r}^{\left( 0 \right)} &=& 0, 
\end{eqnarray}
and $i$-th $\left( i=1,\cdots, h \right)$ state of
\begin{eqnarray}
Q_{k}^{\left( i \right)} = \gamma_{i}^{\frac{1}{2}} V_{ik}
\end{eqnarray}
under any given single composition.  
The important point here is that structure of these $\left( h+1 \right)$ state can be known \textit{a priori} without any information about energy or temperature, i.e., depending only on configurational geometry.
Eq.~\eqref{eq:mod} can be rewritten by a set of energy for such special microscopic state under constant composition:
\begin{widetext}
\begin{eqnarray}
\label{eq:fin}
\Braket{Q_{r}}_{Z} \simeq \Braket{Q_{r}} - 16z^{2}\beta E^{\left( 0 \right)} + 32z^{2}\left( 1-4z \right)\beta^{2} \left( E^{\left( 0 \right)} \right)^{2} + 32z^{3}\beta^{2} \sum_{i=1}^{h} \omega_{i} \left( E^{\left( i \right)} \right)^{2}.
\end{eqnarray}
\end{widetext}
From Eq.~\eqref{eq:fin}, it is now clear that we can predict composition- and temperature-dependence of canonical average from special microscopic states energy at a \textit{single} composition. We finally note that this profound relationship does not generally holds for systems with higher-order correlation (e.g., triplet or quadraplet), since 2-order moment matrix in Eq.~\eqref{eq:moms} is not diagonal at thermodynamic limit. 
In our future work, we should confirm whether or not Eq.~\eqref{eq:fin} can be extended to multicomponent systems, highly desired especially for e.g., high-entropy alloys. 

\section{Conclusions}
By extending our previous approach, we derive a formulation where canonical average for structure at whole compositions in binary systems can be characterized by a set of special microscopic state at a single composition. The results strongly indicate that most information about composition- and temperature-dependence of thermodynamically equilibrium structure for disordered state on multiple compositions can be concentrated to a set of special microscopic state on any single composition derived only from configurational geometry in \textit{non-interacting} system.

\section{Acknowledgement}
This work was supported by Grant-in-Aids for Scientific Research on Innovative Areas on High Entropy Alloys through the grant number JP18H05453 and a Grant-in-Aid for Scientific Research (16K06704) from the MEXT of Japan, Research Grant from Hitachi Metals$\cdot$Materials Science Foundation, and Advanced Low Carbon Technology Research and Development Program of the Japan Science and Technology Agency (JST).

\end{document}